\documentclass[lettersize,journal]{IEEEtran}
\usepackage{amsmath,amsfonts}
\usepackage{algorithmic}
\usepackage{algorithm}
\usepackage{array}
\usepackage[caption=true,font=small]{subfig}
\usepackage{textcomp}
\usepackage{stfloats}
\usepackage{url}
\usepackage{verbatim}
\usepackage{graphicx}
\usepackage{cite}
\usepackage{acronym}
\usepackage{balance}
\usepackage{diagbox}
\usepackage{xcolor}
\usepackage{subcaption}
\usepackage{caption}
\usepackage{colortbl}
\usepackage{tikz}
\usetikzlibrary{shapes.geometric, arrows.meta, positioning}

\acrodef{uav}[UAV]{unmanned aerial vehicle}
\acrodef{fwuav}[FWUAV]{fixed-wing UAV}
\acrodef{hale}[HALE]{high-altitude long endurance UAV}
\acrodef{haps}[HAPS]{high-altitude pseudo-satellite or high-altitude platform station}
\acrodef{sruav}[SRUAV]{single-rotor UAV}
\acrodef{mruav}[MRUAV]{multi-rotor UAV}
\acrodef{huav}[HUAV]{hybrid UAV}
\acrodef{vtol}[VTOL]{vertical takeoff and landing}

\makeatletter
\def\ps@IEEEtitlepagestyle{%
    \def\@oddhead{%
        \parbox[b]{\textwidth}{%
            \centering\scriptsize
            This manuscript has been accepted for publication in
            \textit{IEEE Communications Magazine}.
            This is the author's accepted version and has not been
            fully edited or formatted by IEEE.

            \vspace{2pt}

            \copyright~2026 IEEE. Personal use of this material is permitted.
            Permission from IEEE must be obtained for all other uses,
            in any current or future media, including
            reprinting/republishing this material for advertising
            or promotional purposes, creating new collective works,
            for resale or redistribution to servers or lists,
            or reuse of any copyrighted component of this work
            in other works.
        }%
    }
    \def\@evenhead{\@oddhead}
    \def\@oddfoot{\hfill\thepage\hfill}
    \def\@evenfoot{\hfill\thepage\hfill}
}
\makeatother

\begin{document}

\title{UAV Energy Consumption Models for Wireless Systems Research: Model Selection and Misconceptions}
\author{Mohammadreza Barzegaran and Hamid Jafarkhani
\thanks{Mohammadreza Barzegaran (corresponding author, barzegm1@uci.edu) and Hamid Jafarkhani (hamidj@uci.edu) are with the Department of Electrical Engineering and Computer Science, University of California, Irvine.}}


\maketitle

\begin{abstract}
Unmanned aerial vehicles (UAVs) are gaining increasing attention in wireless systems, providing new opportunities to expand the reach and improve the quality of wireless services. Despite their versatility, UAVs are limited by available energy onboard, which results in significant challenges in deploying UAV-enabled wireless systems. Modeling energy consumption is an essential component of the deployment and trajectory optimization of UAVs. This article presents a practical, application-driven overview of UAV energy consumption models, with a focus on their relevance to wireless systems research. We deliberately exclude data-driven and overly complex models to provide clear and practical guidelines. We categorize the most common types of UAVs, describe their typical flight phases, and review existing energy consumption models and their scopes. We also discuss common mistakes in the use of these models and highlight the existing gaps in the literature. In particular, we show how the use of a wrong model can lead to significant errors in energy consumption calculations. To make the guidance actionable, we provide a compact model summary and a model-selection flowchart that highlights common misconceptions at the steps where they typically occur. Finally, we emphasize the need to develop energy consumption models for missing scenarios.

\end{abstract}

\section*{Introduction}

\Acp{uav} or drones are envisioned as flying wireless platforms 
for coverage extension, aerial base stations and relays, data collection, and rapidly deployable connectivity~\cite{mozaffari2019tutorial,fotouhi2019survey}. This vision is particularly timely for 6G and beyond networks, where ultra-dense deployments, integrated sensing and communication (ISAC), and resilient infrastructure demand adaptive, on-demand network reconfiguration, and \ac{uav}-enabled platforms can dynamically reshape the network topology in 3D space. However, \ac{uav} mobility and operation are constrained by the limited energy onboard, and wireless systems research has increasingly focused on energy-aware trajectory planning and deployment optimization~\cite{zeng2017energy, zeng2019energy,barzegaran2025dynamic,yang2019energy,xiong2022three,kurt2021vision,abdelhady2025optimization}. Neglecting energy constraints results in impractical solutions, while inappropriate models may lead to misleading outcomes.

Existing energy consumption models are derived for a specific type and particular flight scenarios, and are valid only under those scenarios~\cite{zeng2017energy,zeng2019energy,xiong2022three,abdelhady2025optimization,gong2023modeling,Diaz2025}. Applying a model beyond its intended scope can lead to inaccurate or misleading results, which are often difficult to detect. Wireless researchers may not have expertise in flight mechanics, yet they need to employ \acp{uav} as tools to address challenges in wireless systems; a clear explanation of the models and their valid operating conditions remains an unmet need.

To address this need, this article provides a practical and application-driven overview of \ac{uav} energy consumption models tailored for wireless systems research. We identify key factors in selecting an appropriate model and outline common misuses. While aerodynamic information is inherently required in physics-based models, we deliberately exclude overly complex formulations that depend on detailed parameters or are data-driven, and instead focus on models with practical usability for the purpose of deployment and trajectory planning. Specifically, we consider a formulation overly complex when it requires inputs rarely available to wireless researchers, such as airfoil lift and drag polars, blade-element, propeller and motor efficiency maps, or coefficients derived from computational fluid dynamics. Physics-based models are retained when their inputs reduce to a few parameters obtainable from datasheets, such as mass, speed, and a few lumped aerodynamic constants. Data-driven models can be highly accurate for the specific airframe and conditions represented in their training data, and are increasingly important in \ac{uav} digital twins, edge intelligence, real-time control, and hardware-specific energy prediction; however, they lack the generality and analytical tractability required by optimization-based design and are therefore excluded.
This article does not propose new analytical models or derive formulations; it avoids technical derivations and focuses on intuitive explanations, practical guidance, and pointers to the most relevant references.

In the remainder of this article, we distinguish between common \ac{uav} types and their typical flight phases, review the existing energy consumption models and the scenarios for which they are applicable. To the best of our knowledge for the first time, we highlight common misconceptions regarding their use~\cite{mozaffari2019tutorial,fotouhi2019survey}. To make the guidance actionable, we also provide a compact summary of the reviewed models (Table~\ref{tab:modelsummary}) and a model-selection flowchart with integrated misconception warnings (Figure~\ref{fig:flowchart}). We also identify scenarios for which appropriate energy consumption models do not exist and conclude the article.

\section*{UAV types and flight phases}

\begin{figure*}[!t]
\centering
\includegraphics[width=0.98\textwidth]{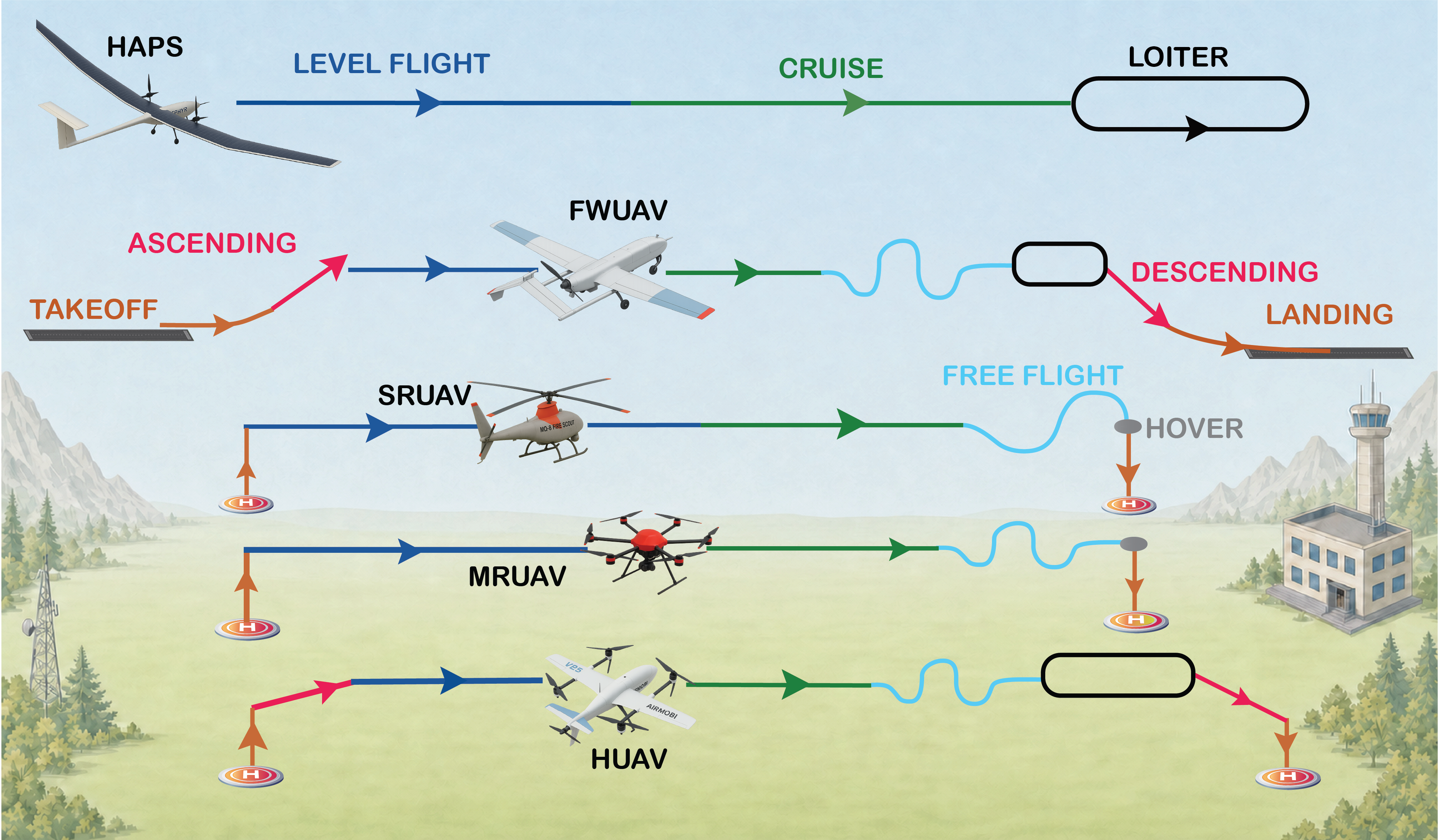}
\caption{High-Altitude Platform Station (HAPS), Fixed-wing UAV (FWUAV), Single-rotor UAV (SRUAV), Multi-rotor UAV (MRUAV), Hybrid UAV (HUAV) are the main types of UAVs considered in wireless systems research. Typical flight phases are takeoff and landing (orange), ascending and descending (red), level flight (blue), cruise (green), free flight (cyan), loiter (black), and hover (gray).}
\label{fig:configurations}
\end{figure*}

This section presents typical \ac{uav} types, classified by airframe characteristics, and their associated flight phases, as commonly considered in wireless systems research. Figure~\ref{fig:configurations} illustrates the main types along with the relevant flight phases and Table~\ref{tab:specsrange} reports specifications of representative \acp{uav} across different types that are used commonly in wireless systems research. The \acp{uav} and reported values in Table~\ref{tab:specsrange} differ in size, power source, and operating conditions; therefore, the table illustrates heterogeneity rather than a controlled performance comparison.

\begin{table}[!t]
\centering
\caption{Representative specifications of different UAVs used in wireless systems research. Endurance, weight, speed, and altitude entries retain the source-specific definitions. No representative HUAV specifications were available in the cited wireless-systems sources.}
\scriptsize
\setlength{\tabcolsep}{2.2pt}
\resizebox{\columnwidth}{!}{%
\begin{tabular}{|l|c|c|c|c|}
\hline
 &
\shortstack{\textbf{HAPS}\\Zephyr S\\\cite{kurt2021vision}} &
\shortstack{\textbf{FWUAV}\\Parrot Disco\\\cite{fotouhi2019survey}} &
\shortstack{\textbf{SRUAV}\\Scout B-330\\\cite{fotouhi2019survey}} &
\shortstack{\textbf{MRUAV}\\DJI S900\\\cite{fotouhi2019survey}} \\ \hline
Endurance & $>$ 2 months & 45 min & 180 min & 18 min\\ \hline
Weight (kg) & 75 & 0.75 & 140 & 4.7\\ \hline
Speed (m/s) & 18 & 22 & 28 & 16\\ \hline
Altitude (m) & 25,000 & 2,500 & 2,000 & 500\\ \hline
\end{tabular}}
\label{tab:specsrange}
\end{table}

\subsection*{Fixed wing UAVs}

\Acp{fwuav} use rigid, non-flapping wings to generate lift and remain airborne through forward motion, similar to conventional airplanes. Owing to their aerodynamic efficiency, they can remain airborne for several hours, carry low to moderate payloads, operate at high altitudes, and achieve higher speeds, outperforming other \ac{uav} types in both range and flight efficiency, see Table~\ref{tab:specsrange} for specifications of a typical mid-size battery-powered \ac{fwuav}. However, they require a minimum forward speed to maintain lift, thus cannot hover; instead, they can loiter by flying in circles. Their climb and descent rates are also constrained by limitations in the wings' angle of attack.\footnote{The angle of attack is the angle between the chord line of a wing and the direction of the incoming airflow. It plays a key role in determining the lift forces and cannot exceed a limit.}

A typical flight profile of a \ac{fwuav} begins with the takeoff phase, which ends when the generated lift is sufficient to achieve and maintain levitation. This phase is often ignored in wireless systems research, as studies in other domains have shown its energy consumption to be small for typical long-endurance missions. However, this assumption is scenario-dependent: the takeoff energy share grows for short missions and depends on the launch method (e.g., hand, runway, or catapult launch), the airframe, and the required climb profile. Once in the air, \acp{fwuav} enter the ascending phase, gaining altitude through forward motion with a positive pitch angle. Due to aerodynamic constraints, the pitch angle and consequently the climb rate are limited. In this phase, \ac{fwuav}'s motion is confined to two dimensions, with a relatively constant vertical speed. After reaching the desired altitude, \acp{fwuav} switch to the level flight phase, maintaining a constant altitude with possibly variable speed. For a constant speed, the phase is called cruise, the most energy-efficient phase of flight. \Acp{fwuav} may also perform 3D maneuvers such as turning, changing altitude, termed free flight. Although it is called ``free,'' the rate of motion change is still limited, e.g., turning rate is typically limited.

In the context of wireless systems, the loiter phase is particularly relevant: \acp{fwuav} fly in circular patterns at a fixed altitude with a bank angle (sideways tilt) to maintain line-of-sight coverage over a designated area. The bank angle is coupled to speed and loiter radius rather than being a universal value, as illustrated in the Motivation section using the model of~\cite{zeng2017energy}. Although the bank angle affects the antenna orientation and sensors' view, it is often ignored in wireless systems research. Finally, \acp{fwuav} enter the descending phase, which mirrors the ascending phase; the descent rate is similarly limited. For a typical mid-size battery-powered \ac{fwuav}, vertical speed does not exceed $4$~m/s. The mission ends with the landing phase, which, like takeoff, is commonly excluded from wireless systems research due to its typically small contribution in long-endurance missions; as with takeoff, this is a scenario-dependent assumption.

\subsubsection*{High-Altitude Platform Stations}
A subclass of \acp{fwuav}, known as \acp{haps} or, less commonly, \acp{hale}, differs significantly from typical \acp{fwuav}. As shown in Table~\ref{tab:specsrange}, they operate at much higher altitudes, often above $20$~km, and can achieve endurances ranging from weeks to months~\cite{kurt2021vision}. To maximize efficiency, they feature an extremely high aspect ratio\footnote{Aspect ratio is the ratio of the wing's length to its width} larger than $30$, which limits maneuverability~\cite{kurt2021vision}. Typically \acp{haps} cannot stop mid-air, many cannot take off or land on their own and must be launched via carrier aircraft or balloon, and some cannot ascend or descend independently due to their mechanically constrained long wings. Thus, we consider that their flight profile is limited to level, cruise, and loiter flights.

In wireless systems research, \acp{haps} are receiving increasing attention and are often envisioned as low-altitude satellites due to their similar operation. This specific \ac{uav} type is expected to attract growing research interest, particularly in the context of 6G development~\cite{kurt2021vision}.

\subsection*{Rotary-wing UAVs}

Rotary-wing \acp{uav} are another class of \acp{uav} that are widely considered in wireless systems research. The primary characteristic is the ability to perform \ac{vtol}. Rotary-wing \acp{uav} are generally classified into two subclasses: \acp{sruav} and \acp{mruav}.

\subsubsection*{Single-rotor UAVs}
The first subclass of rotary-wing \acp{uav}, known as \acp{sruav}, shares the structure, aero-mechanical principles, and flight profile of conventional helicopters, with rotors functioning as a lifting surface. They carry moderate payloads, operate at low to moderate altitudes, and cruise at moderate speeds, Table~\ref{tab:specsrange}. \Acp{sruav} maneuver by tilting the rotor disk to redirect thrust, allowing forward, backward, and lateral motions without changing the body orientation. Their primary advantage is the ability to hover, which makes them well-suited for communication and sensing applications.

\begin{table*}[t]
\centering
\renewcommand{\arraystretch}{1.4}
\caption{Example energy consumption models in the literature and their applicable scopes. Legend: \textcolor{blue}{IG} = energy consumption typically negligible in wireless systems research; a scenario-dependent assumption, \textcolor{gray}{N/A}= Not Applicable, \textcolor{red}{UX}= no tractable model suitable for wireless-systems research is currently available in the literature.}
\begin{tabular}{|c|*{9}{c|}} 
\hline
\diagbox[width=12em, height=3em]{\textbf{UAV Types}}{\textbf{Flight Phases}} 
&Takeoff &Ascending &Level &Cruise &Free &Loiter &Hover &Descending &Landing \\ \hline
Fixed-Wing     &\textcolor{blue}{\texttt{IG}} &\cite{xiong2022three} &\cite{zeng2017energy} &\cite{zeng2017energy} &\cite{xiong2022three} &\cite{zeng2017energy} &\textcolor{gray}{\texttt{N/A}} &\cite{xiong2022three} &\textcolor{blue}{\texttt{IG}} \\ \hline
HAPS           &\textcolor{gray}{\texttt{N/A}} &\textcolor{gray}{\texttt{N/A}} &\cite{zeng2017energy} &\cite{zeng2017energy} &\textcolor{gray}{\texttt{N/A}} &\cite{zeng2017energy} &\textcolor{gray}{\texttt{N/A}} &\textcolor{gray}{\texttt{N/A}} &\textcolor{gray}{\texttt{N/A}} \\ \hline
Single-Rotor   &\textcolor{red}{\texttt{UX}} &\textcolor{gray}{\texttt{N/A}} &\textcolor{red}{\texttt{UX}} &\cite{abdelhady2025optimization} &\textcolor{red}{\texttt{UX}} &\textcolor{gray}{\texttt{N/A}} &\cite{zeng2019energy} &\textcolor{gray}{\texttt{N/A}} &\textcolor{red}{\texttt{UX}} \\ \hline
Multi-Rotor    &\textcolor{red}{\texttt{UX}} &\textcolor{gray}{\texttt{N/A}} &\textcolor{red}{\texttt{UX}} &\cite{gong2023modeling} &\textcolor{red}{\texttt{UX}} &\textcolor{gray}{\texttt{N/A}} &\cite{gong2023modeling} &\textcolor{gray}{\texttt{N/A}} &\textcolor{red}{\texttt{UX}} \\ \hline
Hybrid         &\textcolor{blue}{\texttt{IG}} &\cite{xiong2022three} &\cite{zeng2017energy} &\cite{zeng2017energy} &\cite{xiong2022three} &\cite{zeng2017energy} &\cite{gong2023modeling}  &\cite{xiong2022three} &\textcolor{blue}{\texttt{IG}} \\ \hline
\end{tabular}
\label{tab:energymodels}
\end{table*}

A typical \ac{sruav} flight profile begins with the takeoff phase, a vertical ascent from the ground whose energy consumption can be significant, particularly as the target altitude increases. It is followed by the level flight phase (constant altitude, variable speed); when both altitude and speed are constant, the flight is in the cruise phase. Agile maneuvers involving changes in altitude, speed, and lateral motion are classified as the free flight phase; they are energy-intensive and restricted by the rotor disk tilt limits. During a hover phase, the \ac{sruav} maintains a stationary position and orientation mid-air. The landing phase mirrors takeoff and is similarly excluded in wireless systems research.

For wireless systems research, the level flight and hover phases are particularly relevant. Unlike \acp{fwuav}, which must remain in motion during loitering, \acp{sruav} can hover while preserving their orientation, an advantage for directional antennas and stabilized sensing. However, hovering is significantly more energy-intensive than loitering in \acp{fwuav}.

\subsubsection*{Multi-rotor UAVs}

The second subclass of rotary-wing \acp{uav}, known as \acp{mruav}, has gained growing interest in wireless systems research. Due to the lack of a clear distinction in the literature, the broader term ``rotary-wing \acp{uav}'' is often used when the focus is specifically on \acp{mruav}. This ambiguity has led to incorrect model selection, such as applying \ac{sruav} energy consumption models to \acp{mruav}. Unlike \acp{sruav}, \acp{mruav} carry smaller payloads, operate at lower altitudes, cruise at slower speeds, and exhibit shorter endurance due to a less efficient aerodynamic design (Table~\ref{tab:specsrange}). They follow a flight profile similar to that of \acp{sruav}, with the same considerations in all phases.

\Acp{mruav} replace the complex rotor tilt mechanism of \acp{sruav} with multiple fixed-pitch rotors, reducing costs and improving accessibility for research. Maneuvering is achieved by independently adjusting each rotor's thrust, allowing rapid ascents, descents, lateral movement, and quick orientation changes. However, this agility, comes at the cost of aerodynamic efficiency, leading to lower endurance and payload capacity.

\subsection*{Hybrid UAVs}

\Acp{huav} combine the advantages of both \acp{fwuav} and \acp{mruav}, offering \ac{vtol} capability alongside the aerodynamic efficiency of fixed-wing flight. They ascend and descend vertically like \acp{mruav}, without runways or launch systems, and switch to the fixed-wing mode in the air, significantly extending endurance, range, and cruising speed. Their payload capacity generally lies between that of \acp{mruav} and \acp{fwuav}, and some designs support hovering. Despite these capabilities, \acp{huav} have received limited attention in wireless systems research, and relatively little work has investigated their potential as wireless platforms.

A typical \ac{huav} flight profile begins with a rotary-wing style takeoff, followed by a fixed-wing style ascending phase, and then the level cruise, and free flight phases with the same definitions as for \acp{fwuav}. Due to the hybrid airframe, \ac{huav} efficiency cannot exceed that of a dedicated \ac{mruav} or \ac{fwuav} in any individual flight phase. The profile continues with a hover phase, if supported, or otherwise a loiter phase, followed by a descending phase and a rotary-wing style landing. These characteristics make \acp{huav} particularly attractive for applications such as emergency coverage and military operations, where runways are unavailable and extended endurance is critical.

\section*{Motivation}

Since vendor specifications typically reflect particular test conditions, and even familiar terms denote different quantities across studies~\cite{fotouhi2019survey,kurt2021vision}, models must be compared under a consistent set of assumptions: payload, speed, wind, trajectory, atmospheric conditions, and energy reserve. An analytical result is meaningful only within its stated scope and baseline, so the airframe, flight phase, motion regime, model inputs, neglected effects, and reference case should each be identified explicitly. The work in~\cite{gong2023modeling} shows that \ac{sruav} models cannot be applied directly to \acp{mruav}; they derive separate \ac{mruav} formulations for individual flight phases, validate them against DJI M210 measurements, and demonstrate that rotor count influences power consumption. Therefore, a model should be selected according to both configuration and flight phase. Coupled quantities must likewise remain mutually consistent: the coordinated level-turn relation in~\cite{zeng2017energy} yields a bank angle of approximately $18.5^\circ$ for the Parrot Disco at its reported maximum speed of $22.2$~m/s and an illustrative loiter radius of $150$~m, whereas a $20^\circ$ bank at the same speed implies a radius of roughly $138$~m. Speed, turn radius, and bank angle must thus be specified jointly, particularly in wireless applications where antenna orientation matters.

\section*{Energy Consumption models}

This section reviews existing energy consumption models and their scopes of applicability with respect to \ac{uav} types and flight phases, along with common misuses of these models. As discussed in the introduction, we focus on analytical models, which offer a tractable framework that can be directly incorporated into optimization problems, and exclude data-driven approaches and models relying on detailed aerodynamic parameters. Table~\ref{tab:energymodels} summarizes the available models; the objective is not to exhaustively review, but to provide practical guidance for model selection. In this table, UX entry does not imply that no related model exists in any field; rather, based on our survey of the wireless systems, no analytical model tractable enough for wireless systems optimization is currently available for that case, although related models may exist in forms requiring detailed aerodynamic or hardware-specific inputs. Table~\ref{tab:modelsummary} summarizes the reviewed models, their required inputs, main assumptions, and typical wireless use cases. Building on Tables~\ref{tab:energymodels} and~\ref{tab:modelsummary}, Figure~\ref{fig:flowchart} then organizes the selection procedure as a flowchart, with the common misconceptions attached as warning boxes to the corresponding steps.

\begin{table*}[!t]
\centering
\renewcommand{\arraystretch}{1.15}
\caption{Summary of practical UAV energy consumption models for wireless systems research.}
\footnotesize
\setlength{\tabcolsep}{4pt}
\begin{tabular}{|>{\raggedright\arraybackslash}p{3.2cm}|>{\raggedright\arraybackslash}p{4.0cm}|>{\raggedright\arraybackslash}p{3.0cm}|>{\raggedright\arraybackslash}p{2.7cm}|>{\raggedright\arraybackslash}p{3.4cm}|}
\hline
\rowcolor{teal!12}
\textbf{Model and scope} & \textbf{Required inputs} & \textbf{Main assumptions} & \textbf{Ignored effects} & \textbf{Example wireless use cases} \\ \hline
\textbf{\cite{zeng2017energy}}\newline FWUAV, HAPS, HUAV\newline--- level, cruise, loiter & Mass; speed; acceleration; lumped aerodynamic constants & Constant altitude; no wind & Vertical dynamics; takeoff and landing & Trajectory optimization for aerial relays and base stations; loitering \\ \hline
\rowcolor{gray!10}
\textbf{\cite{xiong2022three}}\newline FWUAV, HUAV \newline--- free flight, ascending, descending & Mass; speed; acceleration; lumped aerodynamic constants & Smooth, bounded maneuvers; no wind & Stall and control-surface limits & 3D trajectory design; secure data collection \\ \hline
\textbf{\cite{zeng2019energy}}\newline SRUAV \newline--- hover, cruise & Mass; speed; lumped rotor parameters  & Steady level flight; helicopter aerodynamics & Acceleration; vertical motion & Hover-and-communicate; trajectory design \\ \hline
\rowcolor{gray!10}
\textbf{\cite{abdelhady2025optimization}}\newline SRUAV \newline--- cruise & Mass; speed; lumped rotor parameters & Constant speed and altitude & Acceleration; maneuvering & UAV-assisted backscatter communications \\ \hline
\textbf{\cite{gong2023modeling}}\newline MRUAV, HUAV \newline--- cruise, hover & Mass; speed; lumped rotor parameters & Steady flight within each phase & Aggressive maneuvers; wind gusts & Aerial base station placement; IoT data collection \\ \hline
\end{tabular}
\label{tab:modelsummary}
\end{table*}

\begin{figure*}[!t]
\centering
\resizebox{0.98\textwidth}{!}{%
\begin{tikzpicture}[
    font=\scriptsize\bfseries,
    >={Stealth[length=1.8mm]},
    term/.style={rectangle, rounded corners=8pt, draw=green!50!black, fill=green!12, thick, minimum height=5.5mm, inner sep=3.5pt, align=center},
    process/.style={rectangle, draw=blue!60!black, fill=blue!8, thick, align=center, inner sep=3.5pt, rounded corners=1pt},
    decision/.style={diamond, aspect=2.4, draw=orange!85!black, fill=orange!15, thick, align=center, inner sep=1pt},
    model/.style={rectangle, rounded corners=2.5pt, draw=teal!60!black, fill=teal!12, thick, align=center, inner sep=3pt, text width=2.55cm},
    openpb/.style={rectangle, rounded corners=8pt, draw=gray!80, fill=gray!15, thick, align=center, inner sep=3.5pt, text width=2.4cm},
    warn/.style={rectangle, rounded corners=2.5pt, draw=red!75!black, dashed, thick, fill=red!7, align=center, inner sep=2.8pt, text width=2.55cm, font=\scriptsize\bfseries},
    lbl/.style={font=\tiny\bfseries, inner sep=1.5pt, fill=white, fill opacity=0.8, text opacity=1},
    arr/.style={->, thick, draw=black!75},
    bang/.style={font=\tiny\bfseries, text=red!75!black}
]
\node[term] (start) at (0,0) {\textbf{Start}};
\node[process, text width=8.0cm] (p1) at (0,-1.15) {Identify the UAV type and the mission's flight phases (Figure~\ref{fig:configurations}, Table~\ref{tab:specsrange})\\Follow the decision path below which implements the\\(UAV type, flight phase) $\rightarrow$ model mapping of Table~\ref{tab:energymodels}};
\node[decision] (d1) at (0,-2.8) {Takeoff or\\landing?};
\node[decision] (d3) at (0,-4.4) {Altitude\\constant?};
\node[decision] (d5) at (0,-6.0) {Speed\\constant?};
\node[decision] (d6) at (0,-7.5) {Motion\\pattern?};
\node[decision] (d2) at (-4.1,-2.8) {MRUAV/SRUAV?};
\node[openpb, text width=2.2cm] (o2) at (-8,-2.8) {\textbf{Typically negligible (IG)}: scenario-dependent, verify for short missions};
\node[openpb, text width=2.2cm] (o1) at (-4.1,-4.5) {\textbf{No tractable model (UX)}};
\node[decision] (d4) at (4,-4.4) {MRUAV/SRUAV?};
\node[openpb, text width=2.2cm] (o4) at (8,-4.4) {\textbf{No tractable model (UX)}\\(free-flight phase)};
\node[decision] (d7) at (-3.9,-6.0) {MRUAV/SRUAV?};
\node[openpb, text width=2.2cm] (o9) at (-8,-6.0) {\textbf{No tractable model (UX)}};

\node[model] (o5) at (-7.5,-9.3) {\textbf{Level-flight model}\\FW/HAPS/HUAV~\cite{zeng2017energy}};
\node[model] (o6) at (-3.95,-9.3) {\textbf{Hover models}\\ SRUAV~\cite{zeng2019energy}\\MRUAV/HUAV~\cite{gong2023modeling}};
\node[model] (o8) at (-0.8,-9.3) {\textbf{Cruise models}\\ FW/HAPS/HUAV~\cite{zeng2017energy}\\SRUAV~\cite{abdelhady2025optimization}\\MRUAV~\cite{gong2023modeling}};
\node[model] (o7) at (3,-9.3) {\textbf{Loiter model}\\FW/HAPS/HUAV~\cite{zeng2017energy}};
\node[model] (o3) at (7.4,-9.3) {\textbf{Free-flight model}\\FWUAV/HUAV~\cite{xiong2022three}};
\node[warn] (w4) at (-8,-4.5) {\textcolor{red!75!black}{\textbf{!}} Do not ignore the altitude dependence of takeoff or landing};
\node[warn, text width=3.5cm] (w2) at (-7.5,-10.7) {\textcolor{red!75!black}{\textbf{!}} Do not apply cruise models to level flight};
\node[warn, text width=5.5cm] (w1) at (-2.4,-10.7) {\textcolor{red!75!black}{\textbf{!}} Never substitute single-rotor and multi-rotor models for one another};
\node[warn, text width=4cm] (w5) at (3,-10.7) {\textcolor{red!75!black}{\textbf{!}} Do not neglect the impact of bank angle on antenna orientation};
\node[warn, text width=4cm] (w3) at (7.4,-10.7) {\textcolor{red!75!black}{\textbf{!}} Do not approximate free flight as piecewise cruise segments};
\node[warn, text width=3.4cm] (w6) at (6.5,-1.15) {\textcolor{red!75!black}{\textbf{!}} Do not treat a HAPS as an ordinary FWUAV:\\ only level, cruise, and loiter are considered. };
\node[process, fill=violet!8, draw=violet!60!black, text width=18.6cm] (ver) at (0,-12) {Verify the selected model's required inputs and assumptions (Table~\ref{tab:modelsummary})\\ if unavailable or violated, revisit the phase identification or treat the case as an open problem};
\node[term] (end) at (0,-13) {\textbf{Apply the model}};
\draw[arr] (start) -- (p1);
\draw[arr] (p1) -- (d1);
\draw[arr] (d1) -- node[lbl, font=\footnotesize\bfseries, right] {No} (d3);
\draw[arr] (d3) -- node[lbl, font=\footnotesize\bfseries, right] {Yes} (d5);
\draw[arr] (d5) -- node[lbl, font=\footnotesize\bfseries, right] {Yes} (d6);
\draw[arr] (d1) -- node[lbl, font=\footnotesize\bfseries, above] {Yes} (d2);
\draw[arr] (d2) -- node[lbl, font=\footnotesize\bfseries, above] {No} (o2);
\draw[arr] (d2.south) -- node[lbl, font=\footnotesize\bfseries, below left, pos=0.25] {Yes} (o1.north);
\draw[arr] (d3) -- node[lbl, font=\footnotesize\bfseries, above] {No} (d4);
\draw[arr] (d4) -- node[lbl, font=\footnotesize\bfseries, above] {Yes} (o4);
\draw[arr] (d4.south) -- ++(0,-0.75) -| node[lbl, font=\footnotesize\bfseries, right, pos=0.25] {No} (o3.north);
\draw[arr] (d5.west) -- node[lbl, font=\footnotesize\bfseries, above, pos=0.25] {No} (d7.east);
\draw[arr] (d7.south) -- ++(0,-0.35) -| node[lbl, font=\footnotesize\bfseries, above, pos=0.25] {No} (o5.north);
\draw[arr] (d7.west) -- node[lbl, font=\footnotesize\bfseries, above, pos=0.25] {Yes} (o9.east);

\draw[arr] (d6.west) -| node[lbl, font=\footnotesize\bfseries, above, pos=0.25] {hover} (o6.north);
\draw[arr] (d6.east) -| node[lbl, font=\footnotesize\bfseries, above, pos=0.25] {loiter} (o7.north);
\draw[arr] (d6.south) -- ++(0,-0.35) -| node[lbl, font=\footnotesize\bfseries, left, pos=0.5] {forward} (o8.north);
\draw[arr] (o1) -- (w4);  
\draw[arr] (o5) -- (w2);  \draw[arr] (w2) -- (w2 |- ver.north);
\draw[arr] (o6) -- (o6 |- w1.north);
\draw[arr] (o8) -- (o8 |- w1.north);
\draw[arr] (w1) -- (w1 |- ver.north);
\draw[arr] (o7) -- (w5);  \draw[arr] (w5) -- (w5 |- ver.north);
\draw[arr] (o3) -- (w3);  \draw[arr] (w3) -- (w3 |- ver.north);
\draw[arr] (ver) -- (end);
\draw[draw=red!75!black, dashed, thin] (w6) -- (p1.east);
\end{tikzpicture}}
\caption{Model-selection flowchart for UAV energy consumption models in wireless systems research, implementing the UAV type and flight phase mapping to the models of Table~\ref{tab:energymodels}. Orange diamonds are decisions, teal boxes are model outcomes, gray boxes are terminal cases, and the violet bar is the final verification step against Table~\ref{tab:modelsummary}. Red dashed warning boxes highlight common misconceptions at the steps where they typically occur.}
\label{fig:flowchart}
\end{figure*}

\begin{figure}
	\centering%
	\subfloat[Impact of target altitude and vertical speed on takeoff energy: Altitude-independent models underestimate energy in a $4~kg$ MRUAV with $100~Wh$ battery. The UAV has a rotor area of $0.535 m^2$, rotor drag coefficient of~$0.011$, and rotor solidity of~$0.045$.\label{fig:takeoff}]{\includegraphics[width=0.91\columnwidth]{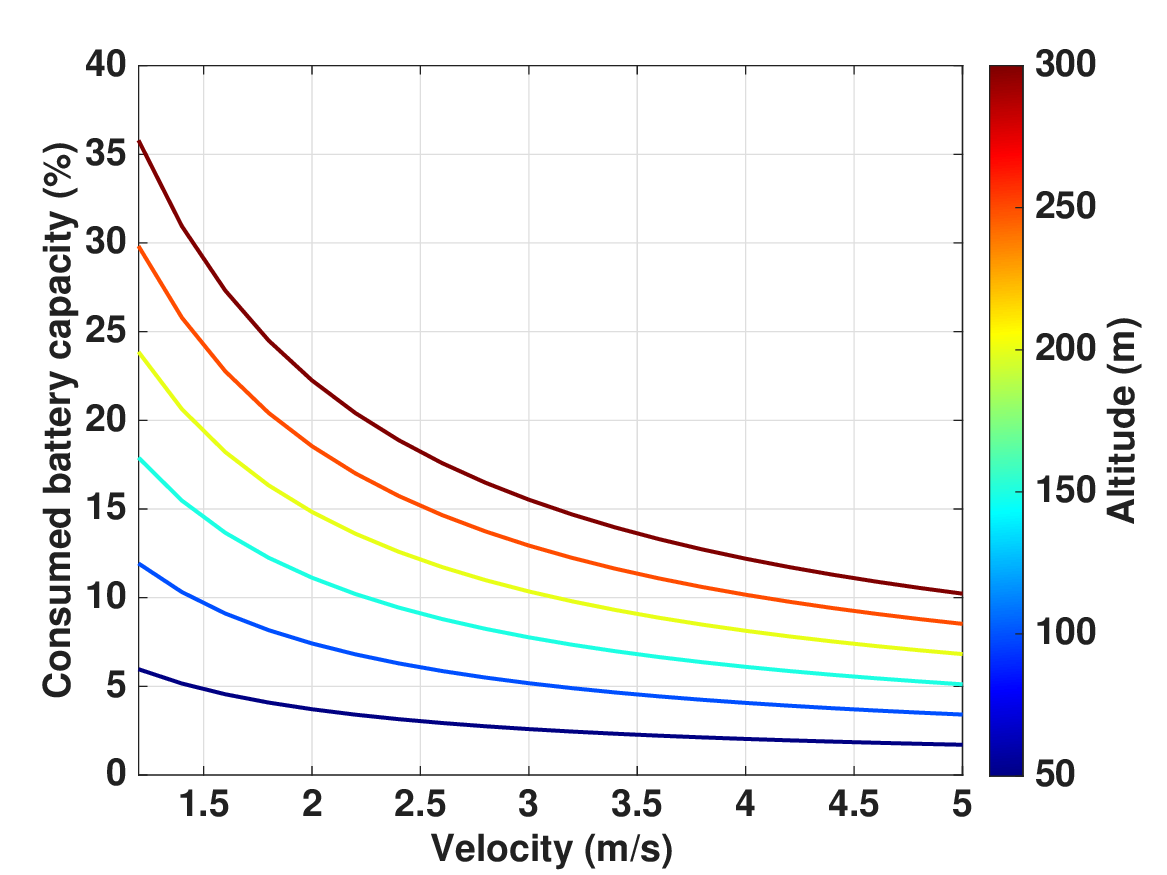}}%
	\\
	\subfloat[Piecewise cruise approximation in level flight phase: Error increases with acceleration and velocity in a $4~kg$ FWUAV with a wing area of $0.15~m^2$, a drag coefficient of $0.012$, and an aspect ratio of $4.5$.\label{fig:levelcruise}]{\includegraphics[width=0.91\columnwidth]{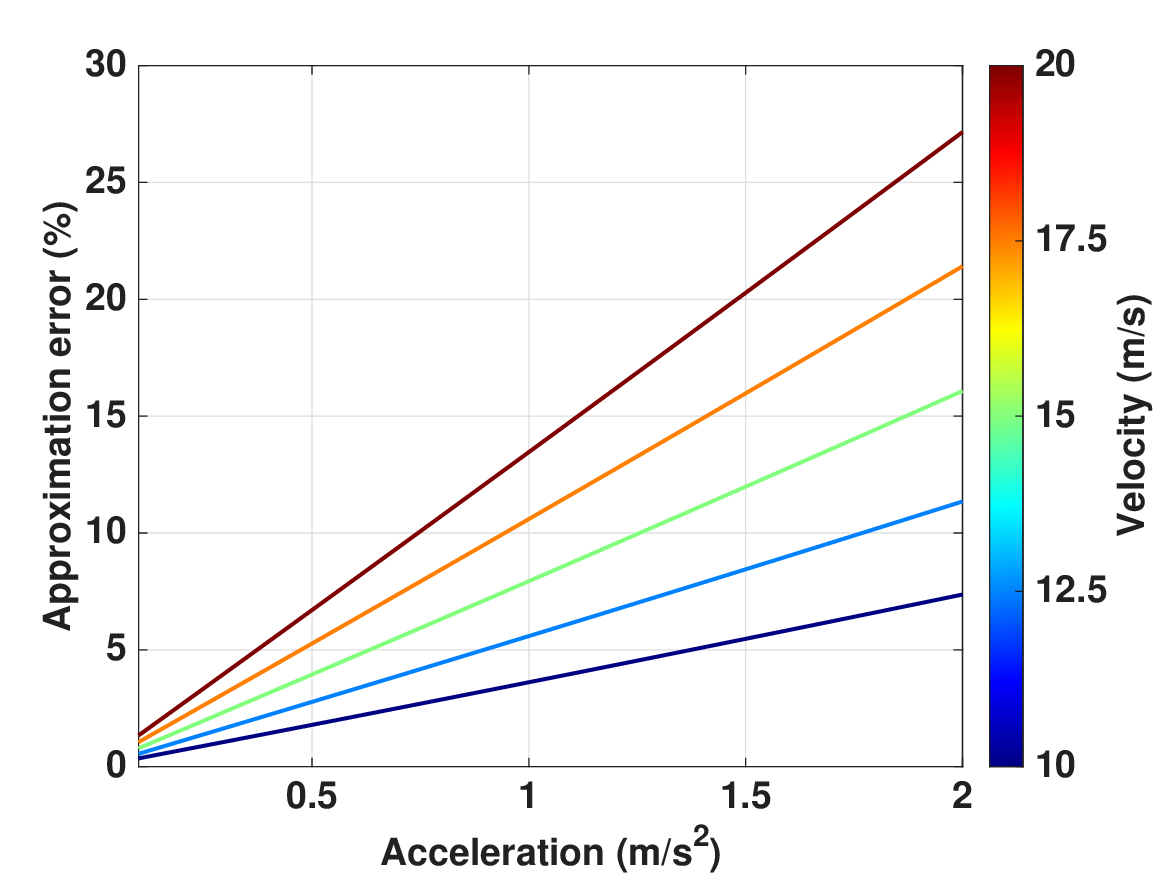}}%
    \\
    \subfloat[Free flight approximation error: Error increases with altitude change when free flight is modeled as piecewise cruise phases in a $4~kg$ FWUAV with a wing area of $0.15~m^2$, a drag coefficient of $0.012$, and an aspect ratio of $4.5$.\label{fig:freecruise}]{\includegraphics[width=0.91\columnwidth]{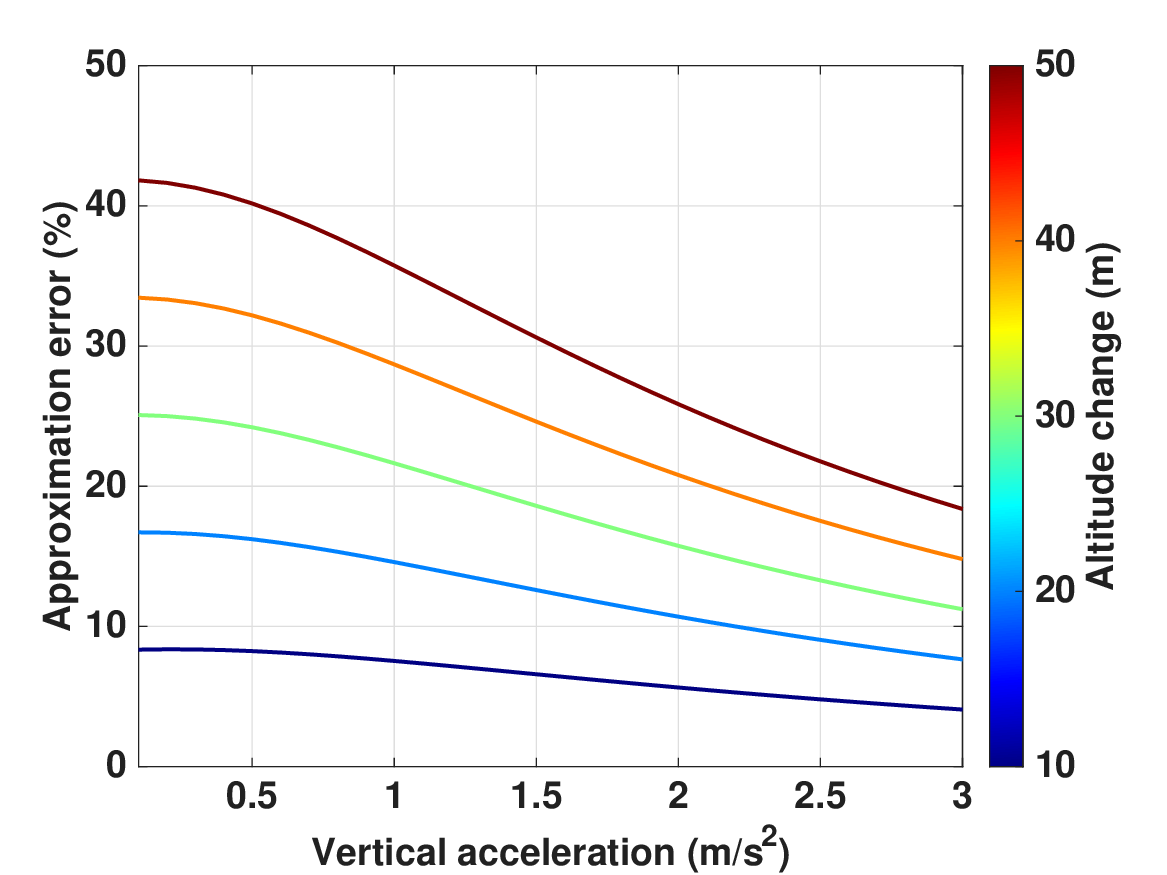}}%
   \caption{Common mistakes in energy consumption models}
\end{figure}

The takeoff phase is usually limited to the initial lift from the ground to a small altitude (e.g., $1$~m). As already discussed, this phase is often overlooked for fixed-wing \acp{uav} (excluding \acp{haps}, for which it is not applicable). For \acp{huav}, the takeoff energy is typically negligible, since altitude is gained during the fixed-wing-style ascending phase; as noted earlier, this assumption is scenario-dependent and should be verified for short missions or demanding climb profiles.

However, for rotary-wing \acp{uav}, the takeoff phase involves not only becoming airborne but also ascending to a desired altitude, and the associated energy consumption is highly dependent on this final altitude. A common misuse is to treat takeoffs to different altitudes in the same way. Refined models accounting for the target altitude exist~\cite{gong2023modeling,yang2019energy}, but they rely on detailed aerodynamic parameters to calculate thrust, limiting their practical applicability. Figure~\ref{fig:takeoff} illustrates the takeoff energy of a typical $4~kg$ \ac{mruav} with $100 Wh$ battery for different target altitudes, based on~\cite{gong2023modeling}. The \ac{mruav} has a rotor disk area of $0.535~m^2$, a drag coefficient of $0.011$, and a rotor solidity of~$0.045$.

Over the range shown, slower ascent increases the battery consumption to about $36\%$. Despite its potential significance, no tractable energy consumption model suitable for wireless-systems research is currently available for the takeoff phase of \acp{sruav} and \acp{mruav}.

Similarly, landing energy is often overlooked, especially for \acp{fwuav} and \acp{huav}, and no tractable model is available in the wireless-systems literature for the landing phase of \acp{sruav} and \acp{mruav}, including the descent from high altitudes.

For the ascending and descending phases of \acp{fwuav} and \acp{huav}, no dedicated tractable energy consumption model is currently available in the wireless-systems literature. A common mistake is to approximate these phases using piecewise level flight phases. Since they involve vertical speed and acceleration components, this introduces large errors. A more accurate approximation is to model them as part of the free flight phase.

As mentioned before, the level flight phase is highly relevant to wireless systems applications and has been well studied, but only for \acp{uav} with fixed wings~\cite{zeng2017energy}. It is not studied for \acp{sruav} and \acp{mruav}, a significant gap, since level flight is the primary phase for sensing and communication applications where these \ac{uav} types are widely used. On the other hand, the cruise phase has been well studied for all \ac{uav} types~\cite{zeng2017energy,zeng2019energy,gong2023modeling}. A common mistake is the use of \ac{sruav} cruise models instead of \acp{mruav} ones, and vice versa. Studies in other communities have shown that these two types exhibit different energy consumption behaviors~\cite{bangura2017thrust, michel2022modeling}. Another common mistake is approximating the level flight as a series of piecewise cruise phases with speed adjusted at each interval, which ignores the energy required for acceleration to adjust the speed. Figure~\ref{fig:levelcruise} compares this approximation using the level flight (baseline) and cruise models in~\cite{zeng2017energy}. The acceleration is tangential in the level flight. The figure shows the resulting underestimation for a $4~kg$ battery-powered \ac{fwuav} with a wing area of $0.15~m^2$, a drag coefficient of $0.012$, and an aspect ratio of $4.5$, using the model in~\cite{zeng2017energy}; the error grows with the required acceleration and is about $8\%$ at a velocity of $15$~m/s and an acceleration of $1$~m/s$^2$.

In realistic wireless applications, \ac{uav} must satisfy domain (obstacles and restricted airspace) constraints and collision constraints, requiring complex maneuvers with acceleration changes, which are classified as the free flight phase. Energy consumption models for the free flight phase have been studied for \acp{fwuav} and \acp{huav}. For \acp{sruav} and \acp{mruav}, limited models exist under uniform conditions, i.e., fixed speed and zero acceleration~\cite{gong2023modeling,gong2024energy,ding20203d}. No tractable general free flight model for these types is currently available in the wireless systems literature, although thrust-based models are available when detailed aerodynamic information is provided~\cite{gong2023modeling}.

For \acp{fwuav}, a common mistake is approximating free flight as a series of piecewise cruise flights, adjusting altitude and velocity in each segment. Since both altitude and speed vary, the resulting error can exceed that of the level flight case. Figure~\ref{fig:freecruise} compares energy consumption of piecewise cruise approximation evaluated with the model of~\cite{zeng2017energy} against the free flight model (baseline) of~\cite{xiong2022three}. The horizontal axis is the vertical acceleration component; the two horizontal acceleration components are fixed at zero. The figure shows the underestimation for a $4$~kg mid-size battery-powered \ac{fwuav} with a wing area of $0.15~m^2$, a drag coefficient of $0.012$, and an aspect ratio of $4.5$. The approximation is only valid when altitude variations are small and vertical speed is nearly constant: with a vertical acceleration of $1~m/s^2$, the error is about $8\%$ for a $10~m$ altitude change and about $36\%$ for a $50~m$ change.

The loiter and hover phases are among the most critical phases in wireless systems research, as they maintain line-of-sight coverage over a designated area. The loiter phase applies only to \acp{fwuav}, \acp{haps}, and \ac{huav}, and is well studied~\cite{zeng2017energy,xiong2022three}. Hover models for \acp{sruav} and \acp{mruav} are also available~\cite{zeng2019energy,gong2023modeling}. A common mistake is to use \ac{sruav} hover models instead of \acp{mruav} ones, despite differences in aerodynamic efficiency that lead to significant estimation errors~\cite{bangura2017thrust, michel2022modeling}.

\section*{Implications for Wireless System Design}
Energy-model errors propagate directly into wireless design metrics such as feasible mission duration, the number of \acp{uav} required for continuous service, coverage and outage probability, link availability, sensing continuity, and network lifetime. For a \ac{mruav} aerial base station, applying the \ac{sruav} hover model of~\cite{zeng2019energy} instead of the \ac{mruav} model of~\cite{gong2023modeling} estimates the feasible service time per charge and, with it, the number of \acp{uav} needed for continuous coverage incorrectly. For a \ac{fwuav} loitering for coverage, the loiter model of~\cite{zeng2017energy} applies; neglecting the bank angle misrepresents the antenna orientation, yielding optimistic link availability and coverage estimates. For a \ac{haps}, treating the platform as an ordinary \ac{fwuav} ignores its limited maneuverability and may produce trajectories that it cannot fly. In a multi-\ac{uav} IoT data-collection mission, ignoring the altitude dependence of takeoff energy (Figure~\ref{fig:takeoff}) can result in wasting a substantial fraction of the battery, causing aborted missions and reduced network lifetime. Finally, for a \ac{huav} in emergency communications, each phase must be matched to the corresponding fixed-wing or rotary-wing model; a single model applied across all phases accumulates errors precisely where reliability matters most.

Furthermore, onboard propulsion energy is only one component of the end-to-end energy efficiency of a \ac{uav}-enabled wireless system, whose platforms must exchange traffic with terrestrial base stations, edge nodes, and the core network. Hence, energy-aware \ac{uav} deployment should be co-designed with an energy-efficient and flexible x-haul/fronthaul transport infrastructure. In this context, optical transport can provide high-capacity, low-latency, and energy-efficient mobile x-haul support for 6G networks.

\section*{Conclusion}
This article presented an overview of UAV energy consumption models for wireless systems research, covering major UAV types, their flight phases, and the applicability of existing models. We revealed two major issues: (\emph{i}) models for certain UAV types and flight phases remain unavailable, and (\emph{ii}) existing models are sometimes applied outside their intended scope, and we quantified the resulting errors. The main lessons learned are: identify the UAV type and flight phase before selecting a model; verify that the model's assumptions, particularly constant speed and altitude, match the scenario; never substitute single-rotor and multi-rotor models for one another; and account for the altitude dependence of takeoff energy in rotary-wing missions. Future research may focus on developing lightweight, physics-based energy consumption models for missing cases, on the role of data-driven models in UAV digital twins and edge intelligence, and on the co-design of energy-aware aerial platforms with energy-efficient and flexible transport infrastructure.

\section*{Acknowledgments}
This work was supported in part by the NSF Award ECCS-2513060.

\bibliography{ref}

@article{mozaffari2019tutorial,
  title={A tutorial on {UAVs} for wireless networks: Applications, challenges, and open problems},
  author={Mozaffari, Mohammad and Saad, Walid and Bennis, Mehdi and Nam, Young-Han and Debbah, M{\'e}rouane},
  journal={IEEE Communications Surveys \& Tutorials},
  volume={21},
  number={3},
  pages={2334--2360},
  year={2019},
  publisher={IEEE}
}

@article{zeng2017energy,
  title={Energy-efficient {UAV} communication with trajectory optimization},
  author={Zeng, Yong and Zhang, Rui},
  journal={IEEE Transactions on Wireless Communications},
  volume={16},
  number={6},
  pages={3747--3760},
  year={2017},
  publisher={IEEE}
}

@article{zeng2019energy,
  title={Energy minimization for wireless communication with rotary-wing {UAV}},
  author={Zeng, Yong and Xu, Jie and Zhang, Rui},
  journal={IEEE Transactions on Wireless Communications},
  volume={18},
  number={4},
  pages={2329--2345},
  year={2019},
  publisher={IEEE}
}

@article{xiong2022three,
  title={Three-dimensional trajectory design for unmanned aerial vehicle-based secure and energy-efficient data collection},
  author={Xiong, Xinxuan and Sun, Chao and Ni, Wei and Wang, Xin},
  journal={IEEE Transactions on Vehicular Technology},
  volume={72},
  number={1},
  pages={664--678},
  year={2022},
  publisher={IEEE}
}

@article{kurt2021vision,
  title={A vision and framework for the high altitude platform station {(HAPS)} networks of the future},
  author={Kurt, Gunes Karabulut and Khoshkholgh, Mohammad G and Alfattani, Safwan and Ibrahim, Ahmed and Darwish, Tasneem SJ and Alam, Md Sahabul and Yanikomeroglu, Halim and Yongacoglu, Abbas},
  journal={IEEE Communications Surveys \& Tutorials},
  volume={23},
  number={2},
  pages={729--779},
  year={2021},
  publisher={IEEE}
}

@article{gong2023modeling,
  title={Modeling power consumptions for multirotor {UAVs}},
  author={Gong, Hao and Huang, Baoqi and Jia, Bing and Dai, Hansu},
  journal={IEEE Transactions on Aerospace and Electronic Systems},
  volume={59},
  number={6},
  pages={7409--7422},
  year={2023},
  publisher={IEEE}
}

@article{yang2019energy,
  title={Energy efficient {UAV} communication with energy harvesting},
  author={Yang, Zhaohui and Xu, Wei and Shikh-Bahaei, Mohammad},
  journal={IEEE Transactions on Vehicular Technology},
  volume={69},
  number={2},
  pages={1913--1927},
  year={2019},
  publisher={IEEE}
}

@article{gong2024energy,
  title={Energy-efficient {3-D} {UAV} ground node accessing using the minimum number of {UAVs}},
  author={Gong, Hao and Huang, Baoqi and Jia, Bing},
  journal={IEEE Transactions on Mobile Computing},
  year={2024},
  publisher={IEEE}
}

@article{ding20203d,
  title={{3D} {UAV} trajectory design and frequency band allocation for energy-efficient and fair communication: A deep reinforcement learning approach},
  author={Ding, Ruijin and Gao, Feifei and Shen, Xuemin Sherman},
  journal={IEEE Transactions on Wireless Communications},
  volume={19},
  number={12},
  pages={7796--7809},
  year={2020},
  publisher={IEEE}
}

@article{bangura2017thrust,
  title={Thrust control for multirotor aerial vehicles},
  author={Bangura, Moses and Mahony, Robert},
  journal={IEEE Transactions on Robotics},
  volume={33},
  number={2},
  pages={390--405},
  year={2017},
  publisher={IEEE}
}

@article{michel2022modeling,
  title={Modeling and validation of electric multirotor unmanned aerial vehicle system energy dynamics},
  author={Michel, Nicolas and Wei, Peng and Kong, Zhaodan and Sinha, Anish Kumar and Lin, Xinfan},
  journal={eTransportation},
  volume={12},
  pages={100173},
  year={2022},
  publisher={Elsevier}
}

@ARTICLE{Diaz2025,
  author={Diaz-Vilor, Carles and Barzegaran, Mohammadreza and Jafarkhani, Hamid},
  journal={IEEE Transactions on Wireless Communications}, 
  title={{Multi-UAV} Energy-Efficient Wildfire Coverage Optimization}, 
  year={2025},
  volume={24},
  number={10},
  pages={8633-8648}}

@article{barzegaran2025dynamic,
  title={Dynamic Deployment of Heterogeneous Wireless Sensor Drone Networks With Limited Communication Range},
  author={Barzegaran, Mohammadreza and Jafarkhani, Hamid},
  journal={IEEE Transactions on Vehicular Technology},
  year={2026},
  volume={75},
  number={1},
  pages={548-563},
  publisher={IEEE}
}

@article{abdelhady2025optimization,
  title={Optimization of Hybrid Laser-Battery-Powered {UAV}-Assisted Backscatter Communications},
  author={Abdelhady, Amr M and Diaz-Vilor, Carles and Barzegaran, Mohammadreza and Jafarkhani, Hamid and Eltawil, Ahmed M},
  journal={IEEE Transactions on Communications},
  year={2025},
  volume={73},
  number={11},
  pages={11690-11706},
  publisher={IEEE}
}

@article{fotouhi2019survey,
  title={Survey on {UAV} cellular communications: Practical aspects, standardization advancements, regulation, and security challenges},
  author={Fotouhi, Azade and Qiang, Haoran and Ding, Ming and Hassan, Mahbub and Galati-Giordano, Lorenzo and Garc{\'i}a-Rodr{\'i}guez, Adrian and Yuan, Jinhong},
  journal={IEEE Communications Surveys \& Tutorials},
  volume={21},
  number={4},
  pages={3417--3442},
  year={2019},
  doi={10.1109/COMST.2019.2906228},
  publisher={IEEE}
}
\bibliographystyle{ieeetr}

\section*{Biographies}
\begin{IEEEbiographynophoto}
Mohammadreza Barzegaran is a Postdoctoral Scholar at the University of California, Irvine. He received the Ph.D. degree in Computer Science from the Technical University of Denmark (DTU), Denmark, in 2021, where he was awarded the Marie Skłodowska-Curie Fellowship. Following his Ph.D., he held a postdoctoral position at DTU as part of a European training network program. Dr. Barzegaran received the M.S. degree in Control Engineering from the University of Tehran, Iran, and the B.S. degree in Aerospace Engineering from Amirkabir University of Technology, Iran. His research interests include wireless and wired sensor networks, real-time and safety-critical cyber-physical systems, and Fog/Edge computing platforms.
\end{IEEEbiographynophoto}

\begin{IEEEbiographynophoto}
Hamid Jafarkhani [F] is a Distinguished Professor at EECS Department, University of California, Irvine. Among his awards are the IEEE Marconi Prize Paper Award in Wireless Communications, the IEEE Communications Society Award for Advances in Communication, the IEEE Wireless Communications Technical Committee Recognition Award, the IEEE Signal Processing and Computing for Communications Technical Recognition Award, the IEEE Eric E. Sumner Award, and the Marconi Society's Marconi Award for Innovation in Wireless Communications. He is the 2017 Innovation Hall of Fame Inductee at the University of Maryland's School of Engineering. He is listed as an ISI highly cited researcher. According to Thomson Scientific, he is one of the top 10 most-cited researchers in the field of ``computer science'' during 1997-2007. He is a Fellow of AAAS, Society of Cardiovascular Magnetic Resonance (SCMR), and National Academy of Inventors (NAI). He is the author of the book ``Space-Time Coding: Theory and Practice.''

\end{IEEEbiographynophoto}

\balance

\end{document}